# First-principles investigation of structural, mechanical, vibrational, thermal, electronic and optical properties of $Na_3Bi$: a topological Dirac semimetal

Maisha Fahmida[1,2], Syed Shovon Mahbub Mahin[1], Suptajoy Barua[1], Ishtiaque M. Syed*[1]

[1]Materials Physics Laboratory, Department of Physics, University of Dhaka, Dhaka-1000, Bangladesh

[2]Department of Physics, Florida State University, Tallahassee, Fl 32304

*Corresponding author e-mail: imsyed@du.ac.bd

## Abstract

Exploring the properties of topological materials has remained an active area of research in condensed matter physics for the past few decades. Here, we present a first-principles investigation of the structural, elasto-mechanical, thermal, electronic, and optical properties of $Na_3Bi$, one of the earliest discovered topological Dirac semimetals, using density functional theory. The hexagonal phase of $Na_3Bi$ is found to be both mechanically and dynamically stable, exhibiting moderate elastic anisotropy with a universal anisotropy index of 0.832. The compound displays a moderate machinability index of 2.386 alongside a low hardness value of 2.005, consistent with its predominantly brittle, covalently bonded character accompanied by a slight metallic contribution. Its comparatively low Debye temperature of 181.39 K points to a mechanically soft lattice with loosely bound atoms. Electronic structure calculations confirm semimetallic behavior, with well-defined Dirac points pinned at the Fermi level and a Fermi surface composed of point-like pockets. Optical analysis across a wide spectral range reveals non-selective, moderate reflectivity of 40–45% and, in the visible region, an unusually high refractive index reaching between 2 and 4.5. $Na_3Bi$ is additionally found to be an efficient absorber of ultraviolet radiation. Together, these findings establish a comprehensive baseline for the mechanical, thermal, and optical behavior of hexagonal $Na_3Bi$ and point to its potential relevance for acoustic damping, UV-sensing, and optoelectronic applications

## Highlights

- First comprehensive DFT study of hexagonal $Na_3Bi$'s elastic and thermal properties.
- $Na_3Bi$ is mechanically and dynamically stable, with moderate elastic anisotropy.
- Low Debye temperature (181 K) reveals a soft lattice with loosely bound atoms.
- Band structure and Fermi surface confirm robust Dirac points at the Fermi level.
- High refractive index and strong UV absorption support optoelectronic applications.

## 1. Introduction

Topological materials are revolutionizing technology with their exceptional properties. These materials possess unique surface properties that facilitate the smooth and efficient flow of electric currents, even in the presence of impurities or defects, which make them highly suitable for advanced electronics, enabling the development of faster and more reliable transistors and sensors [1,2]. These materials are broadly classified in two categories: topological insulators and topological semimetals.

Topological Dirac semimetals are materials where the conduction and valence bands intersect at distinct points in momentum space, protected by time-reversal and inversion symmetries. These band touching points are isolated points with fourfold degeneracy, known as Dirac points or nodes. Low-energy behavior of these points is governed by the 3D Dirac equation [3,4]. Dirac semimetals exhibit novel transport properties in magnetic fields, including ultrahigh electron mobility and large magnetoresistance due to suppressed backscattering, making them promising compounds for high-speed transistors and electronic devices [5]. Scientists have also discovered distinctive orbital texture at the Dirac points, together with a crystal structural phase transition. This indicates symmetry-protected topological superconductivity featuring Majorana fermions [6]. These discoveries may be of interest for topological quantum computing.

The Theoretical prediction of a Dirac semimetal was first studied in 2012 in β-cristobalite $BiO_2$, which hosts three Dirac points at the Fermi level and can be thought of as a three-dimensional counterpart of graphene [3]. Shortly afterward, Z. Wang *et al.* revealed that compounds $A_3Bi$ (A= Na, K, or Rb) can exhibit the presence of extended 3D Dirac points [7]. $Na_3Bi$ is the first 3D Dirac semimetal discovered experimentally using angle-resolved photoemission spectroscopy (ARPES) [7]. ARPES measurements by Xu et al. revealed presence of bulk Dirac states and van Hove singularity in $Na_3Bi$, and predicted that tuning the chemical potential to the Lifshitz transition point could induce exotic phenomena such as unconventional superconductivity or fractional topological states [8,9]. Subsequent studies by Kumar *et al.* showed the remarkable robustness of Dirac points in $Na_3Bi$ under pressure demonstrating that small structural strains do not significantly alter the features of band structure [10]. Negative longitudinal magnetoresistance in $Na_3Bi$ is experimentally observed providing evidence of chiral anomaly [11]. It has also recently been observed that the magnetoconductivity of $Na_3Bi$ at finite temperature exhibits a quadratic relationship with the magnetic field [12]. Optical conductivity studies revealed a tunable mid-infrared absorption window (5-200 μm) sensitive to temperature and carrier density, suggesting potential applications in infrared electro-optical modulators [13]. These collective findings highlight that $Na_3Bi$ is a unique combination of robust topological electronic properties and tunable optoelectronic characteristics.

Theoretically and experimentally, the topological nature of $Na_3Bi$ has been studied. Previous studies have established $Na_3Bi$ as a prototypical three-dimensional Dirac semimetal and have examined its electronic structure, elastic constants, lattice dynamics, phase stability, and thermoelectric properties (of cubic polymorph) [14–16]. To our knowledge, only a limited number of bulk properties of hexagonal pristine $Na_3Bi$ have been systematically explored so far. In particular, its elastic properties, such as, compliance constants, Cauchy pressure, Kleinman parameter, tetragonal shear modulus, elastic anisotropy factors etc. remain insufficiently investigated in the existing literature. Important indicators relevant to industrial applicability, such as hardness parameters and machinability index, have not been thoroughly examined.

In addition, studies of thermophysical and thermodynamic properties such as acoustic impedance, radiation factor, thermal expansion coefficient, dominant phonon wavelength, specific heat, Helmholtz free energy, entropy (as a function of temperature) etc. are still lacking. There have been conflicting remarks regarding the dynamic stability of hexagonal phase in different works [14,17,18]. Thermoelectric properties such as thermal conductivity, electrical conductivity, power factor, Seebeck coefficient and the thermoelectric figure of merit etc. are not investigated in details for hexagonal polymorph. Such parameters are essential for evaluating the potential of $Na_3Bi$ in thermoelectric and acoustic device applications.

In this work, we also extend the investigation to several optical properties that have not been previously reported. These include the dielectric function, refractive index, photoconductivity, reflectivity, absorption coefficient, and energy loss function, all analyzed as a function of photon energy for different photon polarizations.

Although the electronic and dynamical properties of $Na_3Bi$ have been studied earlier, such as the electronic band structure, density of states, and phonon dispersion, we revisit these aspects for completeness and consistency. In addition, we supplement the existing understanding with further analyses, including charge density distribution and Fermi surface characterization, to provide a more comprehensive picture of the material's physical behavior, chemical bonding and topological nature. We also calculated temperature dependence of some thermodynamic and thermoelectric properties of this compound e.g. heat capacity, free energy, Seebeck coefficient, power factor, figure of merit etc.

## 2. Computational Methodology

We have implemented Density Functional Theory (DFT)-based calculations to derive the properties. Vienna Ab initio Simulation Package (VASP) and CAmbridge Serial Total Energy Package (CASTEP) are used which utilize the DFT formalism [19,20]. Within this formalism, the ground state of the crystalline system is obtained by solving the Kohn–Sham equation [21]. Initially, the exchange–correlation effects were described using both the Local Density Approximation (LDA) and the Generalized Gradient Approximation (GGA) within the Perdew–Burke–Ernzerhof (PBE) scheme to investigate the structural properties [22,23]. The norm-conserving pseudopotential with the Koelling-Harmon relativistic approach was used to model the electron-ion interactions [24]. The crystal structure was fully optimized using the Broyden–Fletcher–Goldfarb–Shanno (BFGS) minimization algorithm to obtain the equilibrium geometry of the system [25].

The valence electronic configurations are $2s^2$ $2p^6$ $3s^1$ for Na and $5d^{10}$ $6s^2$ $6p^3$ for Bi. To achieve a satisfactory degree of total energy convergence, the cut-off energy for the plane wave basis set was set to 750 eV. The Brillouin zone (BZ) was sampled using a 10×10×5 Monkhorst–Pack schemed [26] k-point mesh, corresponding to 42 irreducible k-points. This plane wave cut-off energy and k mesh were chosen based on convergence tests. The geometry optimization was performed with convergence thresholds of $10^{-6}$ eV/atom for energy, 0.05 GPa for maximum stress, 0.03 eV/Å for maximum force, and $10^{-3}$ Å for maximum atomic displacement. The density-mixing electronic minimizer was used to perform self-consistent calculations.

The density of states (DOS) and electronic band structure are significantly influenced by spin–orbit coupling (SOC). Therefore, VASP was employed for the calculation of electronic properties. Although CASTEP was used for structural and mechanical property calculations, it does not provide suitable relativistic pseudopotentials for the present material, making SOC calculations impractical within that framework. We used VASP to calculate electronic properties with and without SOC, to examine its effect on electronic properties. Our results show that SOC does not alter the electronic and topological nature of this compound. Since SOC was found to have negligible influence, optical, vibrational, and thermal properties were investigated without SOC using VASP. We used several VASP-compatible packages, including Phonopy for phonon calculations, iFermi for Fermi-surface analysis, and BoltzTraP2 for transport-property calculations [27–30]. VASP also uses a plane wave basis set, where basis size is controlled by cut-off energy. To describe electron-core interaction, the Projector-augmented wave (PAW) method was used [31–33]. The GGA-PBE scheme was implemented as an exchange-correlation functional. Cut-off energy of plane wave basis set was adjusted to 750 eV. For the self-consistent field (SCF) calculations, the energy convergence was set to $10^{-8}$ eV, and the force convergence threshold was −0.01 eV/Å. As the system is metallic, we included smearing. The sigma value was set to 0.2. For static calculation without SOC a mesh size of 24×24×10 was used and for calculations with SOC, a mesh size of 12×12×5 was used.

For the phonon calculations, we employed VASP implemented density functional perturbation theory (DFPT) [34]. The stress tensor was evaluated while keeping the cell shape and volume fixed. We used Phonopy to extract and analyze force constants and obtain the phonon dispersion relation.

For the BoltzTraP2 calculations, the transport properties were evaluated as a function of temperature. All temperature-dependent transport coefficients were computed with respect to the Fermi energy as reference.

## 3. Results and Analysis

### 3.1. Structural Properties

The $Na_3Bi$ lattice has a hexagonal structure that is characterized by the space group $P6_3/mmc$ (No. 194) [7]. Its unit cell consists of 6 Na atoms and 2 Bi atoms. The Na atoms occupy the Wyckoff positions 2b position $\pm\left(0,0,\frac{1}{4}\right)$ and 4f position $\pm\left(\frac{1}{3},\frac{2}{3},\mathrm{u}\right)$, $\pm\left(\frac{2}{3},\frac{1}{3},\frac{1}{2}+\mathrm{u}\right)$ where u=0.583. The Bi atoms are at 2c position $\pm\left(\frac{1}{3},\frac{2}{3},\frac{1}{4}\right)$ [7].

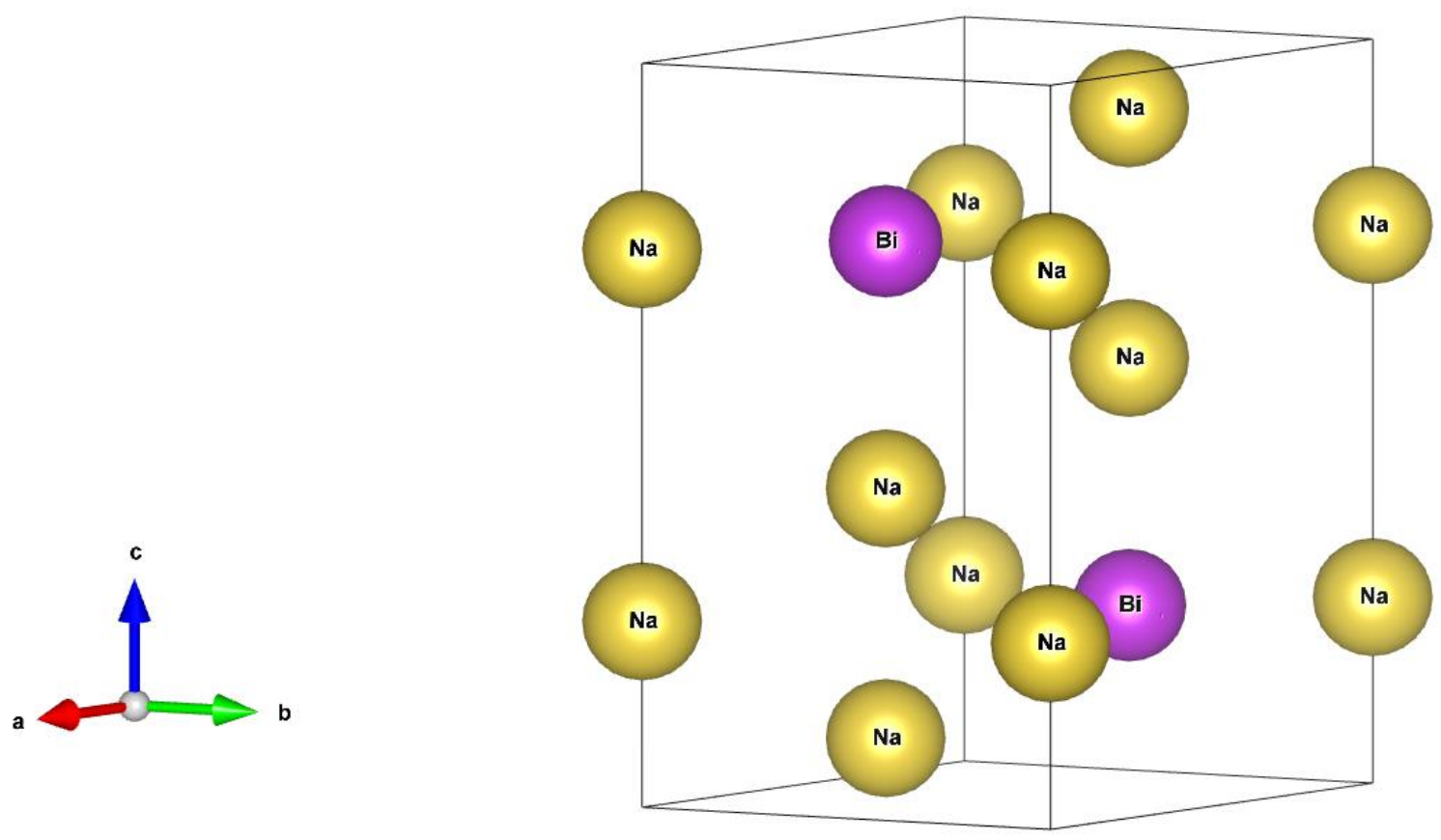

**Figure 1.** Crystal structure of $Na_3Bi$

We calculated the optimized cell volume and lattice parameters of $Na_3Bi$ using GGA and LDA correlation functionals. The results are presented in Table 1, compared with a previously published experimental work [7]. As the GGA findings exhibit more consistency with the experimental values, further calculations are performed using the GGA functional.

**Table 1:** Lattice parameters and cell volumes of $Na_3Bi$ optimized using GGA and LDA functionals (with 42 k-points and 750 eV cut-off energy). The optimized lattice parameters *a*, *b*, and *c* are in Å, the optimized cell volume (*V*) is in Å$^3$ and the error (in %) of volume ($\%\delta V/V$) compared with experimental work [7].

| Functional | a (Å) | b (Å) | c (Å) | V (Å$^3$) | % error |
|---|---|---|---|---|---|
| GGA (this work) | 5.489 | 5.489 | 9.752 | 254.399 | 2.51 |
| LDA (this work) | 5.326 | 5.326 | 9.455 | 232.275 | 6.40 |
| Experimental [7] | 5.448 | 5.448 | 9.655 | 248.17 | |

### 3.2. Elastic Properties

Elastic properties provide essential information about the mechanical stability and dynamical response of a crystal. We present the single-crystal elastic constants and the associated compliance constants calculated using CASTEP in Table 2. Due to symmetry arguments, the hexagonal crystal system has five independent stiffness tensor components which are $C_{11}$, $C_{12}$, $C_{13}$, $C_{33}$, and $C_{44}$ [35]. A crystalline system must satisfy certain conditions to remain mechanically stable under applied stress, known as *Born stability criteria* [36]. For hexagonal systems, the *Born stability criteria* simplify due to the crystal symmetry [35]:

$$C_{11} > |C_{12}|$$
$$2C_{13}^2 < C_{33}(C_{11} + C_{12}) \quad (1)$$
$$C_{44} > 0 \text{ and } C_{66} > 0 \text{ where } C_{66} = (C_{11} - C_{12})/2$$

Stiffness tensor components, also known as elastic constants ($C_{ij}$ in GPa) and compliance constants ($S_{ij}$ in 1/GPa) for the single-crystal system of $Na_3Bi$ are shown below:

**Table 2:** Elastic constants ($C_{ij}$ in GPa) and compliance constants ($S_{ij}$ in 1/GPa) for single crystal $Na_3Bi$.

| **Components, ij** | 11 | 12 | 13 | 33 | 44 |
|---|---|---|---|---|---|
| **C** | 37.222 | 10.091 | 3.759 | 38.917 | 6.911 |
| **S** | 0.02916 | -0.00769 | -0.00207 | 0.02609 | 0.14469 |

It can be seen that all the components of stiffness tensor of $Na_3Bi$ are positive and the stability conditions are satisfied, indicating the mechanical stability of the system. The conclusion of mechanical stability and values of elastic constants are consistent with a previous study [14].

When a material is stretched or compressed in crystallographic *a*, *b,* and *c* directions, $C_{11}$, $C_{22}$, and $C_{33}$ quantify the stiffness respectively. Since the structure is hexagonal, here $C_{11} = C_{22}$. A larger value of $C_{33}$ than $C_{11}$ implies this material's greater compressibility in the *a* and *b* direction than *c* direction. These observations suggest that the intra-planar bonding in the ab-plane is less robust compared to the inter-planar bonding along the out-of-plane direction. The response of a material to shear can be quantified by the components $C_{44}$, $C_{55,}$ and $C_{66}$. For a hexagonal system, $C_{55,}$ and $C_{66}$ are not independent and can be calculated from other elastic constants, [35], where, $C_{44} = C_{55} = 6.911$ GPa and $C_{66} = (C_{11} - C_{12})/2 = 13.566$ GPa. $C_{44}$ is the component that measures the resistance of the crystal to shear deformation in the plane perpendicular to the c-axis, i.e., along the a-axis and b-axis [37]. The comparatively smaller $C_{44}$ elastic constant compared to $C_{66}$ elastic constant suggests an increased susceptibility of $Na_3Bi$ to shear deformation when shear stress is applied along the [010] direction within the (001) plane. $C_{44}$ also provides insights into the hardness of materials. In the case of $Na_3Bi$, the value of $C_{44}$ is smaller compared to $C_{11}$ indicating the material's low resistance to shearing strain. Therefore, the mechanical stability of $Na_3Bi$ is expected to be governed by the shearing strain, rather than the unidirectional strains. The off-diagonal shear components, i.e., $C_{12}$, $C_{13,}$ and $C_{23}$, describe how stress applied to a crystal in one direction can induce strains in other directions. For $Na_3Bi$, the elastic constant $C_{12}$ is smaller than $C_{11}$, indicating a weak resistance to orthogonal distortions.

We calculated the Cauchy pressure ($C''$) for our system (Table 3). The Cauchy pressure, $C''$ is a measurement of the strength of chemical bonding in a material. A positive value of Cauchy pressure indicates strong bonding and ductile nature of the material, whereas a negative value signifies that the material exhibits weak interatomic bonding and is likely brittle [38,39]. Due to symmetry, for the hexagonal system, Cauchy pressure is calculated as $C''_1 = C_{13} - C_{44}$ for {100} and {010} planes, and $C''_2 = C_{12} - C_{66}$ for {001} plane [40].

An additional important parameter for assessing the mechanical stability of a compound is the Kleinman parameter, $\zeta$. It can be calculated using [41]:

$$\zeta = \frac{C_{11} + 8C_{12}}{7C_{11} + 2C_{12}} \tag{2}$$

$\zeta$ is a dimensionless parameter with values ranging from 0 to 1. It describes the relative importance of bond bending and bond stretching. A Kleinman parameter value of 0 implies that the mechanical stability of a compound is governed solely by bond stretching. On the other hand, a value of 1 indicates that bond bending is the dominant factor. Intermediate values between 0 and 1 signify that both bond stretching and bond bending contribute to the mechanical stability [42]. Although originally developed for cubic covalent semiconductors, the Kleinman parameter has since been applied to characterize bond bending/stretching contributions in hexagonal and other lower-symmetry systems [43,44]

Table 3 also contains the tetragonal shear modulus, $C'$. It quantifies the stiffness of a crystal along the direction perpendicular to its two equivalent axes. It can be calculated using the following expression [45]:

$$C' = \frac{C_{11} - C_{12}}{2} \tag{3}$$

**Table 3:** Tetragonal shear modulus ($C'$ in GPa), Cauchy pressure ($C''$ in GPa) and Kleinman parameter ($\zeta$) for $Na_3Bi$.

| $C'$ | $C''_1$ | $C''_2$ | $\zeta$ |
|---|---|---|---|
| 13.57 | -3.15 | -3.47 | 0.42 |

The Cauchy pressure along {100} and {010} planes is -3.15, whereas it is -3.47 along the {001} plane, listed in Table 3. In both directions, the Cauchy pressure is negative for $Na_3Bi$, which suggests that this material is brittle in nature. This value also suggests dominance of directional or covalent bonding in this material.

The Kleinman parameter for $Na_3Bi$ is 0.42. This value indicates that bond stretching contributes slightly more than bond bending to the mechanical strength of $Na_3Bi$.

The tetragonal shear modulus for $Na_3Bi$ is positive, and comparing with several other materials such as $Mo_5PB_2$ (154.89 GPa), $BaGa_2$ (16.25 GPa) and $SrSi_2$ (18.79 GPa) [46–48], we find that it has a lower $C'$, relatively weak resistance to shear-induced structural distortion and is relatively flexible. It is worth noting that, for a hexagonal crystal, the tetragonal shear modulus C′ $(C_{11} - C_{12})/2$ is mathematically identical to the shear elastic constant $C_{66}$ defined earlier; the two are reported separately here only because they are computed via distinct formalisms in the literature, i.e Voigt notation vs. anisotropy analysis, and their near-identical values (13.57 GPa vs. 13.566 GPa) reflect this equivalence rather than two independent calculations.

The polycrystalline moduli such as bulk modulus (B), shear modulus (G) and Young's modulus (Y) can be calculated using the single-crystal elastic constants and compliances. Values of these parameters along with the Poisson's ratio (υ), machinability index ($\mu_M$), and hardness (H) for $Na_3Bi$ are listed in Table 4. To determine the distribution of stress or strain in a polycrystalline aggregate under an external load, two extreme cases can be considered. In the first case, the uniform strain in the aggregate can be equated to the external strain, which is referred to as the Voigt approximation [49]. In the second case, the uniform stress can be equated to the external stress, known as the Reuss approximation [50]. In Hill's approximation, the arithmetic average of the two limiting cases is taken, assuming partial continuity of both stress and strain [51]. This approximation accurately reflects the actual conditions in polycrystalline materials by taking into account energy considerations. The shear modulus describes a material's resistance to shear deformation, whereas the bulk modulus measures its capacity to withstand volumetric changes under hydrostatic pressure. In hexagonal systems, the shear and bulk moduli under the Reuss, Voigt, and Hill approximations are determined using the following expressions [51]:

$$G_R = \frac{15}{4(S_{11}+S_{22}+S_{33}) - 4(S_{12}+S_{13}+S_{23}) + 3(S_{44}+S_{55}+S_{66})} \tag{4}$$

$$G_v = \frac{1}{15}(C_{11}+C_{22}+C_{33}-C_{12}-C_{13}-C_{23}) + \frac{1}{5}(C_{44}+C_{55}+C_{66}) \tag{5}$$

$$B_R = \frac{1}{(S_{11}+S_{22}+S_{33}) + 2(S_{12}+S_{13}+S_{23})} \tag{6}$$

$$B_v = \frac{1}{9}(C_{11}+C_{22}+C_{33}) + \frac{2}{9}(C_{12}+C_{13}+C_{23}) \tag{7}$$

$$G_H = \frac{G_V+G_R}{2} \text{ and } B_H = \frac{B_V+B_R}{2} \tag{8}$$

In Table 4, Young's modulus, $Y$, Poisson's ratio ($\upsilon$), machinability index ($\mu_M$), and hardness ($H$) are also tabulated. These parameters are calculated using the following equations [52]:

$$Y = \frac{9BG}{3B+G} \tag{9}$$

$$\upsilon = \frac{3B-2G}{2(3B+G)} \tag{10}$$

$$\mu_M = \frac{B}{C_{44}} \tag{11}$$

$$H = \frac{(1-2\upsilon)Y}{6(1+\upsilon)} \tag{12}$$

**Table 4:** Bulk modulus (*B* in GPa) and shear modulus (*G* in GPa) using Reuss, Voigt, and Hill's approximations; Hill's approximation for Pugh's ratio (*B/G*), Young's modulus (*Y* in GPa), Poisson's ratio ($\upsilon$), machinability index ($\mu_M$) and hardness parameter (H) for polycrystalline $Na_3Bi$.

| $B_R$ | $B_V$ | $B_H$ | $G_R$ | $G_V$ | $G_H$ | B/G | Y | $\upsilon$ | $\mu_M$ | H |
|---|---|---|---|---|---|---|---|---|---|---|
| 16.465 | 16.509 | 16.487 | 10.174 | 11.861 | 11.018 | 1.496 | 27.031 | 0.227 | 2.386 | 2.005 |

In the case of $Na_3Bi$, the bulk modulus is larger than shear modulus. This means $Na_3Bi$ is more resistant to volume compression than to shear-induced plastic deformation. This result also suggests that mechanical failure in $Na_3Bi$ is dominated by shear effects, in agreement with results from the single-crystal elastic constants. Compared to many other hexagonal compounds [37,53,54], the small elastic moduli of $Na_3Bi$ reflect its soft mechanical nature. The Pugh's ratio (B/G) is a dimensionless quantity that characterizes the ductile or brittle nature of a material. Solids exhibiting a high Pugh's ratio (> 1.75) are more prone to ductility, while solids with a low value of Pugh's ratio (< 1.75) are more prone to brittleness [55]. So, according to our calculated result, $Na_3Bi$ has a brittle character.

Young's modulus, *Y*, defines the material's ability to resist deformation in both tension and compression. Materials with greater Young's modulus values exhibit increased stiffness and need greater stress to produce a certain amount of strain. The small value of *Y* compared to other materials such as $Mo_5PB_2$ (344.63 GPa), $BaGa_2$ (50.62 GPa), $SrSi_2$ (77.82 GPa) and $MoTe_2$ (95.37 GPa) [46–48,56], implies that $Na_3Bi$ has low stiffness, and will elongate a lot when subjected to tensile stress. So, this compound is unable to endure significant tensile stress without breaking. It can be useful in flexible electronics and strain-tunable devices.

Poisson's ratio ($\nu$) defines the relationship between transverse and axial strains in a material under external stress, indicating how much it expands laterally when compressed axially. It generally ranges from −1 to 0.5, with lower values signifying enhanced shear stability. This parameter also provides insight into material ductility or brittleness. Poisson's ratio of less than 0.26 suggest brittleness of the material, whereas a value greater than 0.26 suggests that the material is ductile [57,58]. In solids governed by central interatomic forces or having dominant metallic bonding, Poisson's ratio generally falls within the range of 0.25–0.50. For purely covalent compounds, Poisson's ratio is typically close to 0.10, whereas purely metallic solids exhibit values around 0.33 [59]. Thus, the value of $\upsilon$ for $Na_3Bi$ (0.227) implies that this material is brittle in nature and there is a mixture of metallic and covalent bonding, directional or covalent bond being dominant. This value is consistent with a previous study [14].

The machinability index ($\mu_M$) is defined as a ratio ($B/C_{44}$) and is a dimensionless quantity [60]. where higher values correspond to improved machinability. The high value obtained for $Na_3Bi$ indicates that the material is readily machinable. Along with $\mu_M$, hardness, *H* provides additional insights about machinability. Materials exhibiting high hardness (*H*) are widely used in cutting tools and wear-resistant coatings, despite posing significant challenges during machining [47]. The moderately low value of *H* of $Na_3Bi$ indicates its good machinability, as suggested by the machinability index.

Elastic anisotropic factors and directional bulk moduli offer a comprehensive insight into the anisotropies inherent in bonding and other mechanical features of a system. Additionally, we calculated $\alpha$ and $\beta$, which represent the relative changes in the $b$ and $c$ axes, respectively, as a function of deformation along the $a$ axis. Using elastic theory and the concept of bulk modulus, assuming equal strains perpendicular to the stress direction, we derived an expression for the lower bound of the bulk modulus ($B_{relax}$) [52]:

$$B_{\text{relax}} = \frac{\Lambda}{(1 + \alpha + \beta)^2} \tag{13}$$

where

$$\alpha = \frac{(C_{11} - C_{12})(C_{33} - C_{13}) - (C_{23} - C_{13})(C_{11} - C_{13})}{(C_{33} - C_{13})(C_{22} - C_{12}) - (C_{13} - C_{23})(C_{12} - C_{23})}$$

$$\beta = \frac{(C_{22} - C_{12})(C_{11} - C_{13}) - (C_{11} - C_{12})(C_{23} - C_{12})}{(C_{22} - C_{12})(C_{33} - C_{13}) - (C_{12} - C_{23})(C_{13} - C_{23})}$$

and

$$\Lambda = C_{11} + 2C_{12}\alpha + C_{22}\alpha^2 + 2C_{13}\beta + C_{33}\beta^2 + 2C_{23}\alpha\beta$$

The upper bound ($B_{unrelax}$) is obtained from Eq. (13) by substituting $\alpha = \beta = 1$.

The bulk moduli along the $a$, $b$ and $c$ axes are defined as [52]:

$$B_a = a\frac{dP}{da} = \frac{\Lambda}{1 + \alpha + \beta} \tag{14}$$

$$B_b = b\frac{dP}{db} = \frac{B_a}{\alpha} \tag{15}$$

$$B_c = c\frac{dP}{dc} = \frac{B_a}{\beta} \tag{16}$$

All these linear bulk moduli have been calculated using single crystal elastic constants using the above relations and tabulated in Table 5. For the hexagonal system, $\alpha = 1$.

**Table 5:** The bulk modulus ($B_{relax}$ in GPa) and its upper bound ($B_{unrelax}$ in GPa), bulk modulus along the crystallographic axes a, b, c ($B_a$, $B_b$, $B_c$) and α, β for $Na_3Bi$.

| $B_{relax}$ | $B_{unrelax}$ | $B_a$ | $B_b$ | $B_c$ | α | β |
|---|---|---|---|---|---|---|
| 16.469 | 16.512 | 51.573 | 51.573 | 45.559 | 1 | 1.132 |

The lower value of $B_c$ compared to the values of $B_a$ and $B_b$ indicates $Na_3Bi$ exhibits greater compressibility under stress applied along the $c$-axis. Anisotropy of the material can also be determined from the values of $B_a$, $B_b$ and $B_c$. Since $B_a$=$B_b$, there is no anisotropy along the $ab$-plane but anisotropy is seen along the $c$-direction.

### 3.3. Elastic Anisotropy Analysis

Elastic anisotropy refers to the directional dependence of a material's elastic behavior, meaning its stiffness or compliance changes with orientation. Understanding elastic anisotropy is essential in materials science research and several fields of engineering, where the behavior of materials under different loading conditions and orientations significantly impacts the performance and reliability of structures and components. In this work, we investigated elastic anisotropy using elastic anisotropic indexes, such as the universal elastic anisotropic index ($A^U$), percent compressibility and shear anisotropy ($A_{comp}$ and $A_{shear}$), shear anisotropic factors ($A_1$, $A_2$ and $A_3$) and compressibility anisotropy factors ($A_{Ba}$ and $A_{Bc}$) for $Na_3Bi$. These results are presented in Table 6.

The universal anisotropic index, $A^U$, serves as a general measure of elastic anisotropy in single crystals, incorporating effects from both bulk and shear moduli and is written as [61]:

$$A^U = 5\frac{G_V}{G_R} + \frac{B_V}{B_R} - 6 \geq 0 \tag{17}$$

When $A^U = 0$, the crystal shows isotropic behavior, and the further it deviates from zero, the more pronounced the elastic anisotropy [61].

Shear anisotropic factors quantify the degree of anisotropy in atomic bonding within different crystallographic planes [52]. The factors $A_1$, $A_2$ and $A_3$ correspond to the {100} shear planes between the <011> and <010> directions, {010} shear planes between the <101> and <001> directions, and {001} shear planes between the <110> and <010> directions, respectively, and are defined as [52]:

$$A_1 = \frac{4C_{44}}{C_{11} + C_{33} - 2C_{13}} \tag{18}$$

$$A_2 = \frac{4C_{55}}{C_{22} + C_{33} - 2C_{23}} \tag{19}$$

and

$$A_3 = \frac{4C_{66}}{C_{11} + C_{22} - 2C_{12}} \tag{20}$$

For an isotropic crystal, the shear anisotropic factors $A_1$, $A_2$ and $A_3$ are equal to one. Deviations from unity indicate the extent of elastic anisotropy in the crystal [52]. The bulk modulus anisotropy along the *a* and *c* directions with respect to the *b* direction is given by [52]:

$$A_{B_a} = \frac{B_a}{B_b} = \alpha$$

and,

$$A_{B_c} = \frac{B_c}{B_b} = \frac{\alpha}{\beta} \tag{21}$$

An alternative approach is the percent elastic anisotropy, which quantifies the extent of elastic anisotropy in a crystal. The bulk and shear percent anisotropies, denoted ($A_B$ and $A_G$), are expressed as [52]:

$$A_B = \frac{B_V - B_R}{B_V + B_R}$$

and,

$$A_G = \frac{G_V - G_R}{G_V + G_R} \quad (22)$$

where $B_V$, $B_R$; $G_V$ and $G_R$ are the bulk and shear moduli in the Voigt and Reuss approximations, respectively.

**Table 6:** Shear anisotropic factors $A_1$, $A_2$, $A_3$, universal anisotropic index $A^U$, and bulk and shear percent anisotropies $A_B$ (in %), $A_G$ (in %) and compressibility anisotropy factors $A_{B_a}$, $A_{B_c}$ for $Na_3Bi$.

| $A_1$ | $A_2$ | $A_3$ | $A^U$ | $A_B$ | $A_G$ | $A_{B_a}$ | $A_{B_c}$ |
|---|---|---|---|---|---|---|---|
| 0.403 | 0.403 | 1 | 0.832 | 0.1 | 7.7 | 1 | 0.884 |

Here, $A^U$ is non-zero which suggests anisotropy in $Na_3Bi$, but a lower value compared to other materials suggest low to moderate anisotropy [53,62]. The percent anisotropies are also insignificant, compared to the other anisotropic materials [52,53,56]. The reason behind this is $B_V/B_R$ and $G_V/G_R$ values being close to 1. For $Na_3Bi$, anisotropy in shear moduli (7.7%) is greater than the anisotropy in bulk moduli (0.1%). Also, $A_1$ is equivalent to $A_2$ due to $C_{44}$=$C_{55}$, due to symmetry constraint of hexagonal structure [63]. All these anisotropic indices suggest a subtle anisotropy is present in $Na_3Bi$.

Using ELATE, 2D and 3D visual representation of several elastic parameters such as Young's modulus, linear compressibility, shear modulus and Poisson's ratio is possible [64,65]. Using components of Stiffness matrix, ELATE plots directional dependence of these quantities. Circular shape in 2D and spherical shape in 3D signify isotropic elastic behavior in all directions. Deformation from these shapes suggests anisotropic behavior. The highest and lowest values of these parameters and their ratios which indicate anisotropy are listed below. Figure 2 shows 2D [in (xy)ab, (xz)ac and (yz)bc planes] and 3D dependence of these parameters for $Na_3Bi$. The blue and green curves indicate highest and lowest values of these quantities respectively. Table 7 presents the maximum and minimum values of these parameters, calculated by ELATE.

**Table 7:** Maximum and minimum values of Young's modulus (Y), linear compressibility (β), shear modulus (G), Poisson's ratio (ν) and anisotropy of these parameters.

| **Y** | | | **β** | | | **G** | | | **ν** | | |
|---|---|---|---|---|---|---|---|---|---|---|---|
| $Y_{max}$ (GPa) | $Y_{min}$ (GPa) | $A_Y$ | $\beta_{max}$ ($TPa^{-1}$) | $\beta_{min}$ ($TPa^{-1}$) | $A_\beta$ | $G_{max}$ (GPa) | $G_{min}$ (GPa) | $A_G$ | $\nu_{max}$ | $\nu_{min}$ | $A_\nu$ |
| 38.32 | 20.42 | 1.87 | 21.95 | 19.39 | 1.13 | 16.83 | 6.91 | 2.44 | 0.48 | 0.08 | 6.00 |

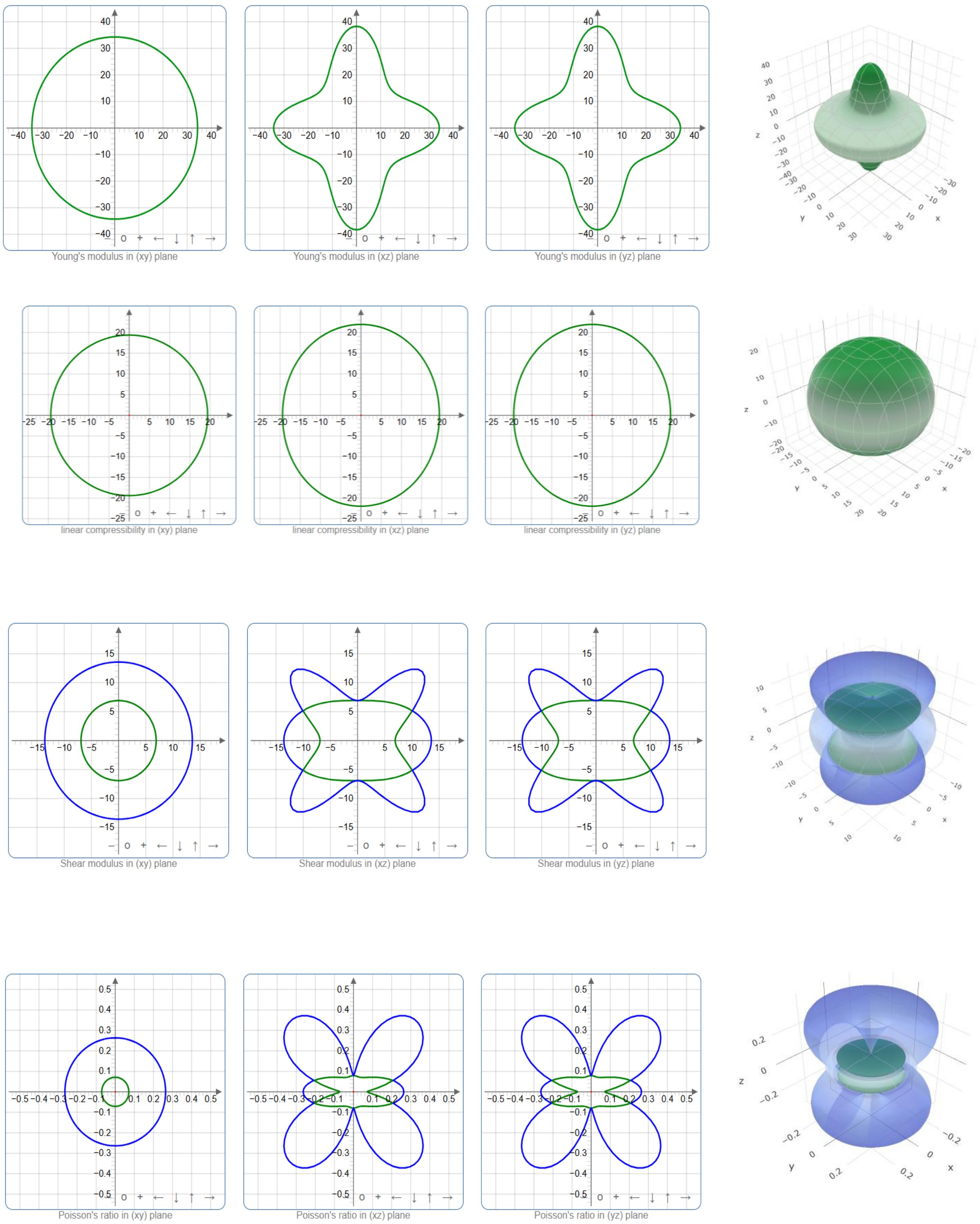

**Figure 2:** Directional dependence of Young's modulus (Y), linear compressibility (β), shear modulus (G) and Poisson's ratio (ν) of $Na_3Bi$.

Young's modulus (*Y*), shear modulus (*G*) and Poisson's ratio (ν) are isotropic in xy plane, but shows moderate anisotropy in yz and xz planes. This is another indication of the presence of subtle elastic anisotropy in $Na_3Bi$. Linear compressibility (β) is almost isotropic in all directions. So, the material has almost similar resistance to compression in all crystallographic directions.

### 3.4. Thermomechanical Properties

Elastic wave behavior is a central topic in multiple disciplines, including geology, materials science, musical instrument design, and medical sciences, since it strongly impacts a material's electrical and thermal conductivities. Using the single-crystal elastic constants, we calculated several thermomechanical properties, such as acoustic impedance (*Z*) and radiation factor. These quantities play a critical role in understanding the material's acoustic response. Acoustic impedance of a material quantifies the extent to which sound energy is transferred between two different media. A higher acoustic impedance results in greater reflection of incident sound energy. It is defined as [66]:

$$Z = \sqrt{\rho G} \tag{23}$$

Where $\rho$ is the density of a material in kg/m$^3$ and $G$ is the shear modulus in Pa. The radiation factor is associated with the intensity of sound radiation and indicates how efficiently a material emits sound waves. Materials with high radiation factors are more efficient at radiating sound waves, while materials with low radiation factors are less efficient. It can be expressed as [66]:

$$I \propto \sqrt{\frac{G}{\rho^3}} \tag{24}$$

where the proportionality constant depends on the geometry of the soundboard and the properties of the surrounding medium. The radiation factor also influences the coupling between the propagation of structural waves in vibrating components and the noise created by the structure. An increased radiation factor signifies more efficient emission of sound energy away from the structure. The computed acoustic impedance and radiation factor for $Na_3Bi$ are shown in Table 8. For understanding the classical and quantum behavior of solids, the Debye temperature ($\theta_D$) is a very crucial parameter. This parameter is defined as the temperature above which the vibrational modes of atoms in a solid lattice are excited to such an extent that quantum mechanical effects become significant. At temperatures higher than the Debye temperature, the solid behaves in the classical way, where the energy of the vibrations is proportional to the temperature. Below the Debye temperature, quantum effects govern lattice vibrations and the vibrational energy becomes quantized. Experimentally, the Debye temperature can be obtained from the elastic constants and average sound velocity through the relation [67]:

$$\theta_D = \frac{h}{k_B}\left[\left(\frac{3n}{4\pi}\right)\frac{N_A\rho}{M}\right]^{\frac{1}{3}} v_m \tag{25}$$

where $h$ is Planck's constant, $k_B$ is Boltzmann's constant, $N_A$ is Avogadro's number, $n$ denotes the number of atoms in the molecule, $\rho$ denotes the density, $M$ denotes the molecular weight, and the mean elastic wave velocity is expressed as $v_m$. The mean elastic wave velocity $v_m$ in the polycrystalline material can be measured by [52,68]:

$$v_m = \left[\frac{1}{3}\left(\frac{2}{v_t^3} + \frac{1}{v_l^3}\right)\right]^{-\frac{1}{3}} \quad (26)$$

where $v_l$ and $v_t$ denotes the longitudinal and transverse elastic wave velocity through the material which can be calculated using [67]:

$$v_l = \sqrt{\frac{B+\frac{4G}{3}}{\rho}} \text{ and } v_t = \sqrt{\frac{G}{\rho}} \quad (27)$$

The Debye temperature ($\theta_D$), logitudinal ($v_l$), transverse ($v_t$) and mean elastic velocity ($v_m$) in the crystal have been calculated for $Na_3Bi$ using the above formulas and are presented in Table 8.

Another important parameter, the Grüneisen parameter, $\gamma$, is a dimensionless quantity that quantifies the extent of lattice anharmonicity. It describes how the phonon frequency changes with a change in the volume of a crystal, providing valuable insights into its thermal and mechanical properties. A large value of $\gamma$ implies a significant degree of anharmonicity, indicating that the atoms in the crystal lattice are engaging in more intricate interactions with one another. The Grüneisen parameter can be calculated from the Poisson's ratio ($v$) using the following formula [69]:

$$\gamma = \frac{3(1+v)}{2(2-3v)} \quad (28)$$

Thermal expansion coefficient is an important parameter which signifies change in dimension with unit change in temperature. It is also a useful indicator of anharmonicity of atomic bonding. Thermal expansion coefficient, $\alpha_{th}$ is also a key design parameter in microelectronics, geophysics, aerospace and turbine engineering etc. It can be calculated using [66]:

$$\alpha_{th} = \frac{\gamma k_B}{B\,\Omega} \quad (29)$$

where, $\Omega$ = Volume occupied by an atom.

Quanta of lattice vibration is known as phonon. At a particular temperature, phonons obey Bose-Einstein statistics [70]. At a particular wavelength, the phonon spectra have a peak, corresponding to maximum energy density. This characteristic wavelength is known as dominant phonon wavelength, $\lambda_d$. It is also a useful parameter which sets the criteria for alloying and nanostructuring to increase thermoelectric figure of merit. It can be calculated using [71]:

$$\lambda_d = \frac{12.566\, v_m}{T} \times 10^{-12} \quad (30)$$

These parameters are also presented in Table 8 below.

**Table 8:** The density ($\rho$ in g/cm$^3$), the acoustic impedance ($Z$ in Rayl) and the radiation factor ($\sqrt{G/\rho^3}$ in m$^4$/kg-s), longitudinal ($v_l$), transverse ($v_t$) and mean ($v_m$) elastic wave velocity (in m/s), Debye temperature ($\theta_D$ in K), Grüneisen parameter ($\gamma$), thermal expansion coefficient ($\alpha_{th}$ in $K^{-1}$ ), and the dominant phonon wavelength in 300 K ($\lambda_d$ in m) of $Na_3Bi$, obtained from the polycrystalline elastic modulus.

| $\rho$ | $Z$ ($\times 10^6$) | $\sqrt{G/\rho^3}$ | $v_l$ | $v_t$ | $v_m$ | $\theta_D$ | $\gamma$ | $\alpha_{th}$ ($\times 10^{-5}$ ) | $\lambda_d$ ($\times 10^{-12}$ ) |
|---|---|---|---|---|---|---|---|---|---|
| 3.63 | 6.32 | 0.48 | 2931.21 | 1742.51 | 1929.36 | 181.39 | 1.40 | 3.69 | 80.81 |

The calculated acoustic impedance of $Na_3Bi$ (Z = 6.32 × $10^6$ Rayl) is much lower than that of common metals such as silver (Z ≈ 2.79 × $10^{10}$ Rayl) and steel (Z ≈ 4.07 × $10^{10}$ Rayl), and is comparable in magnitude to low-density polymeric foams (Z ≈ 1.0 × $10^7$ Rayl) [66]. So, it does not efficiently transmit or radiate sound. Such properties make it suitable for sound absorption or isolation. Therefore, this material has potential use in acoustic linings, damping layers for speakers or musical instruments, and anti-vibration pads [72]. $Na_3Bi$ is a candidate for intermediate matching layer in transducer applications to minimize the impedance mismatch between dissimilar materials and thereby reduce sound reflection at interfaces [73,74].

The Debye temperature for $Na_3Bi$ is 181.39 K which is a relatively low value. A low value of $\theta_D$ indicates that $Na_3Bi$ is soft, consistent with the results of the elastic moduli computations. The values of Debye temperature and sound velocities are consistent with a previous study [14].

The calculated value of the Grüneisen parameter, $\gamma$, for $Na_3Bi$ is 1.4. This value, when compared with other materials [47,62], indicates the presence of an intermediate interaction between atoms, rather than a simple or complicated interaction. The thermal expansion coefficient value of this material is comparable with many topological materials, ($SrSi_2$, $Nb_2P_5$, $Mo_5PB_2$), which are claimed to have moderate to high thermal expansion coefficients, although this value is much lower than $BaGa_2$ [46–48,75]. This result suggests that the material is sensitive to temperature change, making it unsuitable for high temperature applications. Interestingly, dominant phonon wavelength of this material is comparable to $BaGa_2$, while being much less than $SrSi_2$, $Nb_2P_5$ and $Mo_5PB_2$ [46–48,75].

### 3.5. Electronic Properties

#### 3.5.1. Band Structure

The band structure of $Na_3Bi$ was studied both with and without spin-orbit coupling (SOC). We used VASP for these calculations. Figure 3 depicts the GGA-calculated electronic band structure of $Na_3Bi$ using the optimized structure along the high symmetry points inside the Brillouin zone (BZ). Figure 3(a) shows the illustration of the calculated electronic band structure without spin-orbit coupling (SOC), whereas Figure 3(b) shows the band structure with SOC taken into account. The figures have the Fermi level, $E_f$, defined at 0 eV.

For the electronic band structure without SOC, the valence and conduction bands intersect at the Fermi level, indicating no band gap at the Fermi level, $E_f$. The valence bands meet the conduction band only at the Γ points, then dispersed linearly along the 3D momentum space forming the "Dirac cones". The band's touching points are called the "Dirac points". Band structure with SOC shows a single conical electronlike band above the Fermi energy, which qualitatively agrees with experimental results [76,77]. Our calculated band structures with and without SOC are consistent with previous studies [7,8,77]. Di Bernardo *et al.* investigated electronic and topological properties of $Na_3Bi$ in details and found that band structures using GGA, meta-GGA, HSE06 functionals are almost similar. GW approximated band structure is slightly different but doesn't alter the electronic or topological nature (Dirac like crossing) [77]. Due to the coexistence of time reversal and inversion symmetries, each Dirac point has a four-fold degeneracy. This allows for the linearization of band dispersions, resulting in a 3D Dirac semimetal. Upon introducing SOC, the one of the Dirac points is not gapped and the crossing of the linear dispersion is protected by crystal symmetry, retaining electronic nature of $Na_3Bi$ as a semimetal.

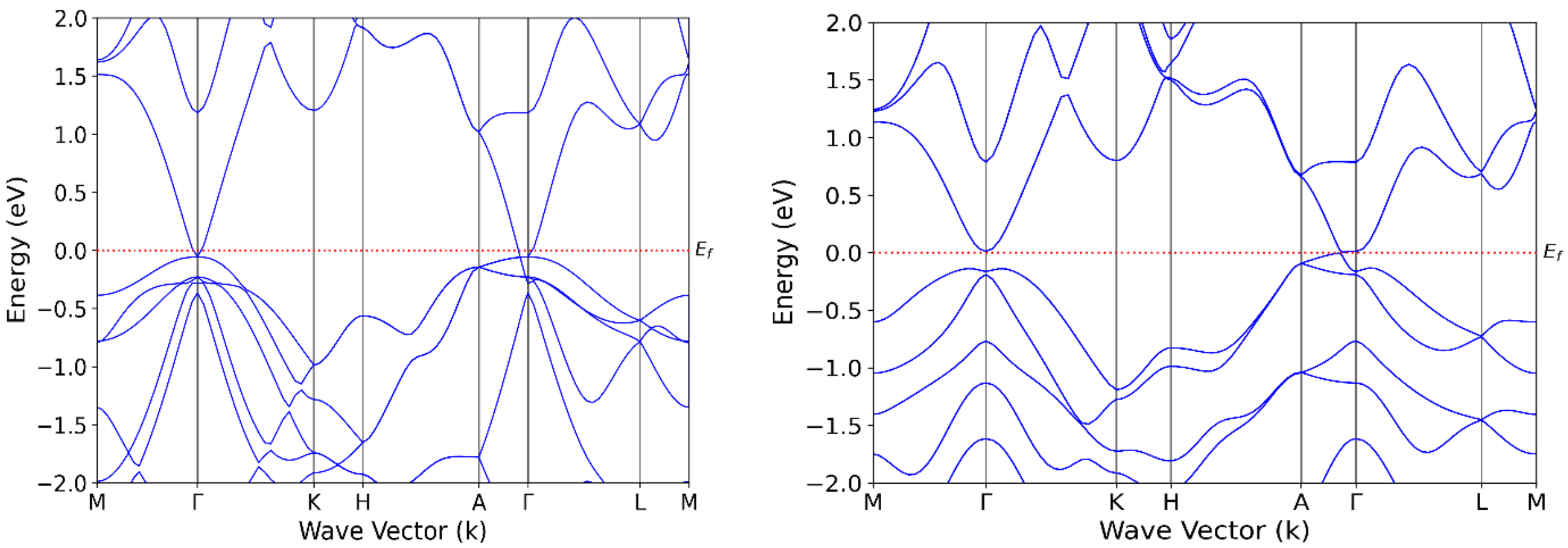

**Figure 3:** The band structure (a) without SOC and (b) with SOC of $Na_3Bi$ along the high symmetry path of BZ.

### 3.5.2. Density of States

Total Density of States (TDOS) as well as the atom projected density of states (PDOS) are calculated in this work. They play an important role in understanding the electronic properties of materials and can be used to calculate the electrical conductivity, optical properties, and other properties of materials. Figure 4 shows the TDOS and PDOS as a function of energy. The Fermi level ($E_f$) is used as a reference and denoted by the vertical dashed line. From Figure 4, it can be seen that there is a non-zero electron density at the Fermi level, $E_f$, which indicates that $Na_3Bi$ has a metallic nature. This result is in good agreement with our electronic band structure calculations and is consistent with a previous study [7].

It can be observed that the Na-3s, Na-2p, and the Bi-6p states are the top contributors of electrons to the TDOS. Figure 4 also suggests that the Bi-6p dominates the valence band below the Fermi level, whereas the Na-3s and Na-2p states account for the majority of the contributions in the conduction bands above the Fermi level. The energy bands between -2.5 eV and the Fermi level are composed of the Na 3s and 2p orbitals, as well as the Bi 6p orbital. This suggests the presence of hybridization between these orbitals. The bands located above the Fermi level mostly originate from the Na 3s and 2p orbitals, which experience hybridization with one another.

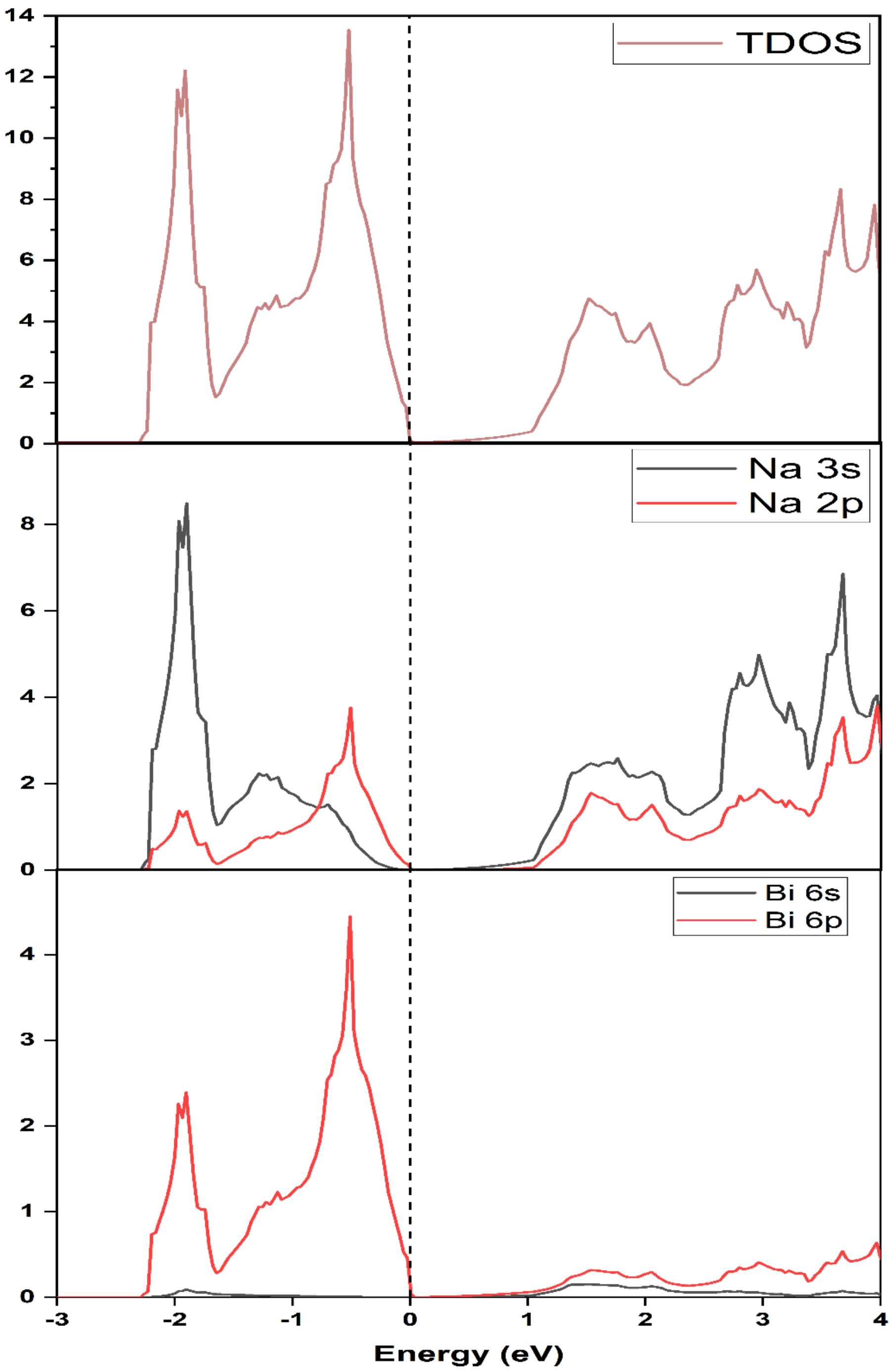

**Figure 4:** TDOS and PDOS for $Na_3Bi$.

### 3.6 Charge Density Distribution

To investigate bonding characteristics in $Na_3Bi$, charge density distribution is plotted and analyzed. In Figure 5, the two-dimensional charge density distribution on the (1 0 0) and (0 0 1) planes are visualized. Due to symmetry (1 0 0) plane has the same charge density distribution as the (0 1 0) plane. The scale shows charge density measurements, where violet indicates low charge density and gradually blue, green and yellow colors suggest higher charge density. Here, the charge density is distributed almost equally among the atoms, which is characteristic of the absence of ionic bonding. The charge density around the atoms is not purely spherical, indicating dominant directional, covalent-like bonding. In addition, the nearly uniform contribution of charge density throughout the crystal, other than near the atoms, suggests the presence of metallic bonding. Overall, the system exhibits a mixture of covalent and metallic bonding, with covalent interactions being dominant. This is consistent with our previous analysis.

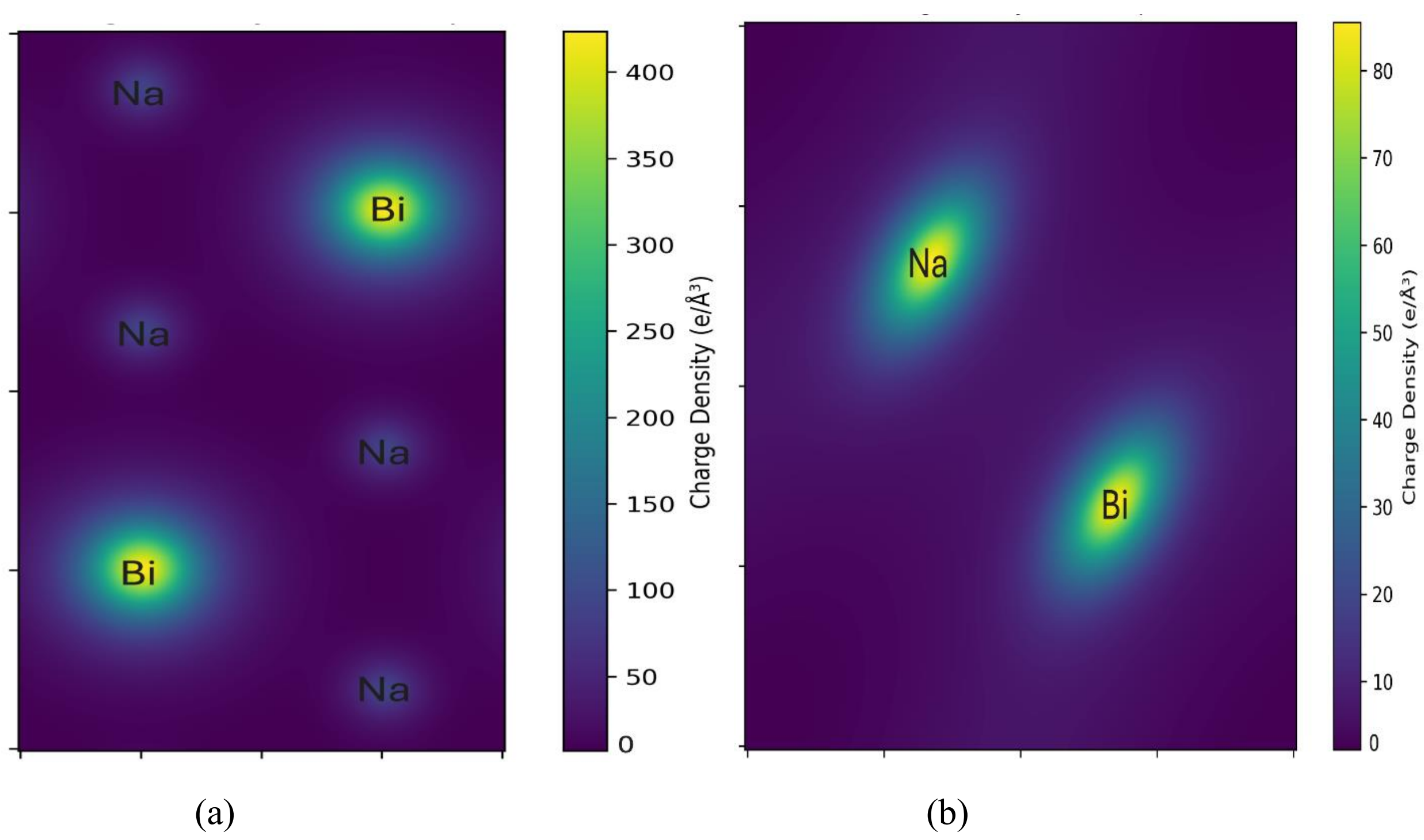

**Figure 5:** Electronic charge density distribution for $Na_3Bi$ in the (a) (1 0 0) and (b) (0 0 1) planes

### 3.7 Fermi Surface

Fermi surface is a ground state property of a system. It is an iso-energy surface separating occupied and unoccupied electronic states in reciprocal space at 0 K. Its topology strongly influences various quantum phenomena, including ferromagnetism, topological insulation, and superconductivity. Additionally, it provides insight into the behavior of occupied and unoccupied states in a metallic system at low temperatures. Figure 6 presents the calculated Fermi surface. Only three bands intersect the Fermi level, in agreement with our computed band structure. The Fermi surface of $Na_3Bi$ is therefore constructed from these three bands. The plot clearly indicates Dirac semimetal nature of our system. The Fermi surfaces are very small, almost showing point like pockets. The shape of Fermi surfaces indicates linear (Dirac like) crossings of bands near Dirac point. The Fermi surfaces also show rotational symmetry and there is electron-hole symmetry. All of these findings are consistent with the conclusion that $Na_3Bi$ is a Dirac semimetal.

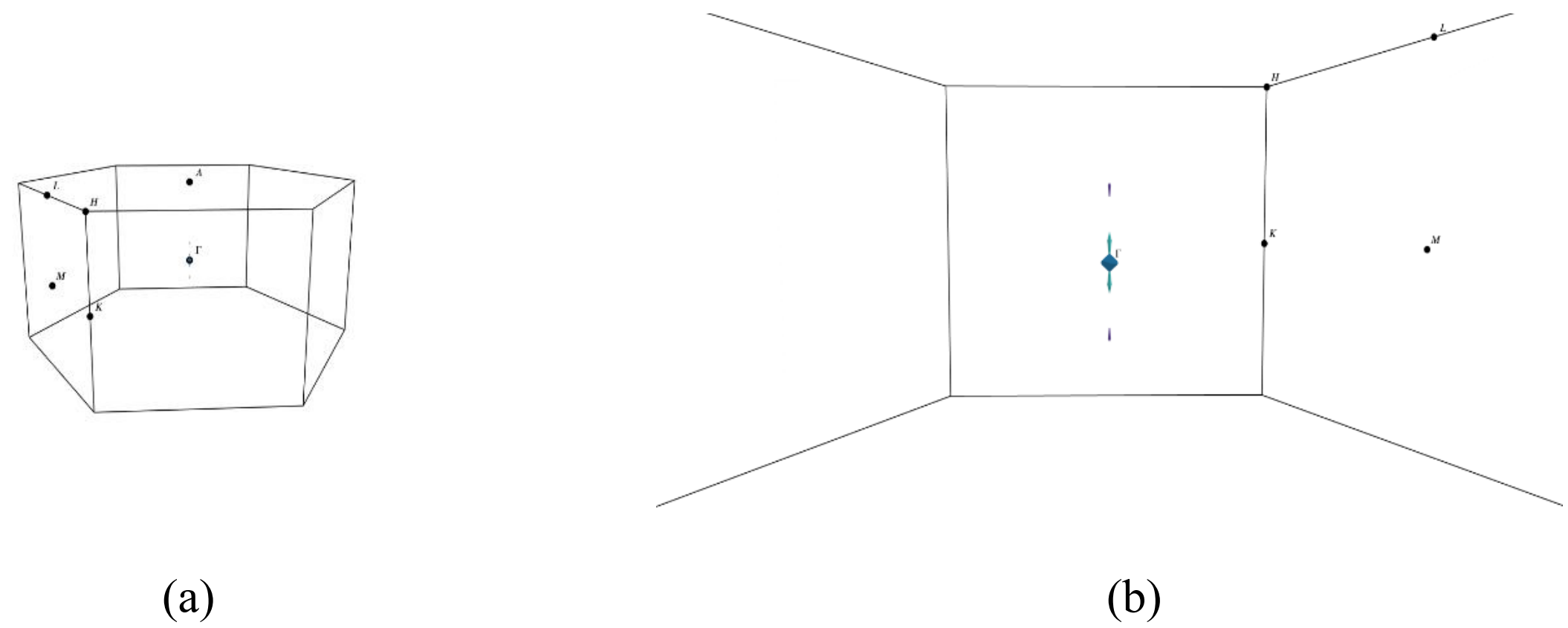

(a) (b)

**Figure 6:** Fermi surface plot of $Na_3Bi$: (a) Distant view (b) Close view

### 3.8 Phonon dispersion relation and density of states

Phonon band structure and density of states have direct and indirect influence on dynamical properties of solids [78]. Phonon band structure reveals a crystal's dynamic lattice stability information, potential phase transitions, as well as its thermal properties, including heat conduction, expansion, and heat capacity [78,79]. Moreover, optical phonons dictate optical properties of a compound. Phonon calculations were carried out using VASP implemented density functional perturbation theory (DFPT) [34,80]. The resulting force constants were processed using Phonopy to obtain the phonon dispersion relations [25,27,28,80].

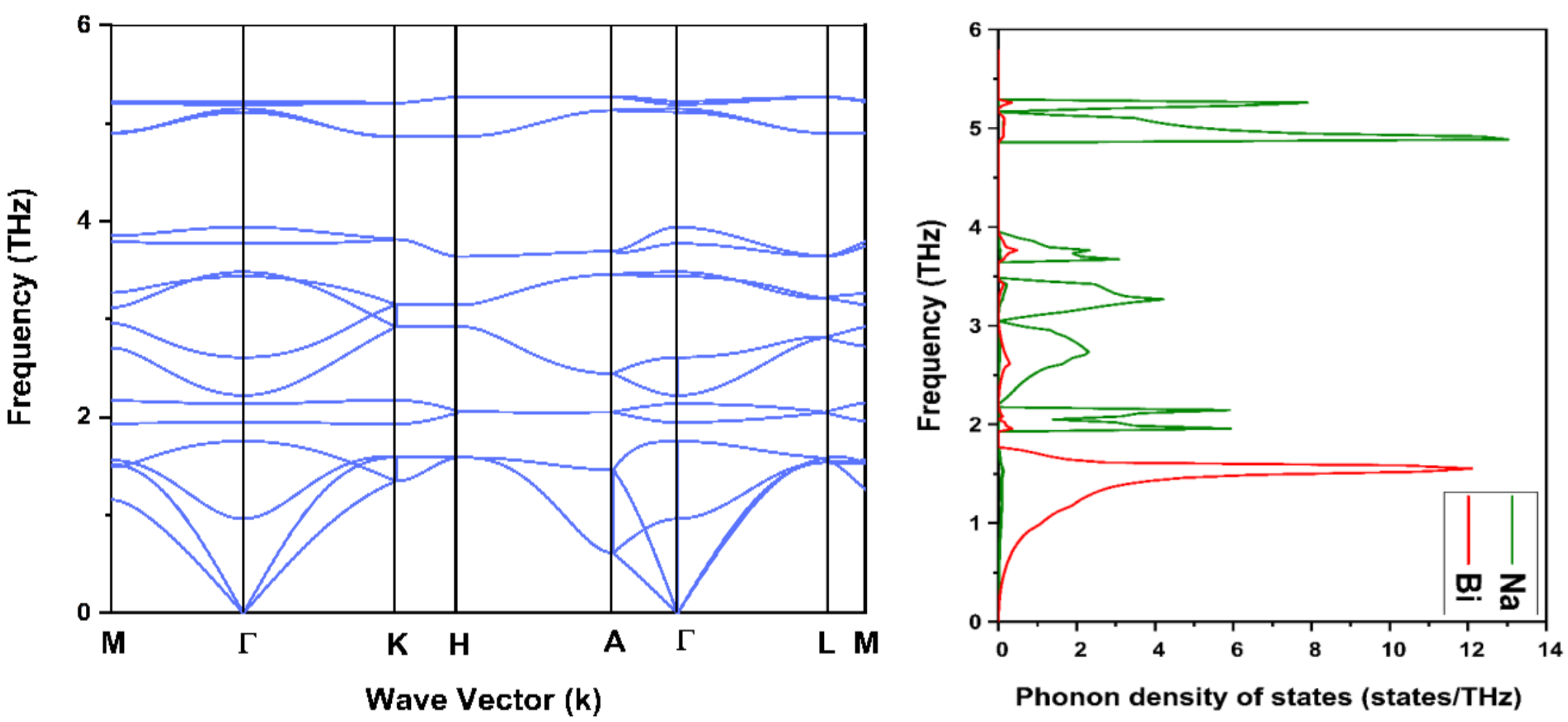

**Figure 7:** Phonon dispersion curve and atom projected phonon density of states for $Na_3Bi$.

The phonon dispersion curve obtained in this work closely resembles that reported by Desai *et al.* [17]A system is dynamically stable if there is no negative frequency in phonon spectra. Imaginary (negative) frequencies indicate soft phonon modes and dynamic instability. $Na_3Bi$ phonon spectra do not have any imaginary frequencies. Therefore, the compound is dynamically stable, consistent with two previous studies [14,17]. In a crystal, the number of phonon modes equals three times the number of atoms in the unit cell. For $Na_3Bi$, with 8 atoms per cell, this yields 24 phonon modes: 3 acoustic and 21 optical. The lowest-frequency modes are acoustic, while the optical modes, which determine optical behavior, are split into upper and lower branches due to the mass difference between Na and Bi. The absence of a phononic bandgap between the acoustic and lower optical branches implies favorable thermal conductivity. We also calculated atom resolved phonon density of states. Lower frequency modes are due to heavy atom Bi and higher frequency modes are due to comparatively lighter element Na.

### 3.9. Thermodynamic and thermoelectric properties

Specific heat is defined as the amount of heat required to raise the temperature of a unit mass of a substance by one degree Celsius. The specific heat at constant volume can be expressed as [70]:

$$C_v = \left(\frac{dE}{dT}\right)_v$$

where, E is the total energy associated with lattice vibration due to finite non zero change in temperature. Under the harmonic approximation, phonon partition function and specific heat can be written as [81]:

$$Z = \prod \frac{\exp(-\hbar\omega/2kT)}{1-\exp(\hbar\omega/kT)} \tag{31}$$

$$C_V = \sum k \left(\frac{\hbar\omega}{kT}\right)^2 \frac{\exp(\hbar\omega/kT)}{(\exp(\hbar\omega/kT)-1)^2} \tag{32}$$

where, the product and sum respectively are over allowed modes. From phonon partition function, Helmholtz free energy and entropy can be written as [28,70]:

$$F = -kT \ln Z \tag{33}$$

$$S = -\left(\frac{dF}{dT}\right) \tag{34}$$

Using the harmonic approximation used in Phonopy [28] code, it is possible to calculate these thermodynamic properties. This code was used to extract the specific heat at constant volume $C_v$, Helmholtz free energy F and entropy S over a temperature interval. All of these calculated results are given below:

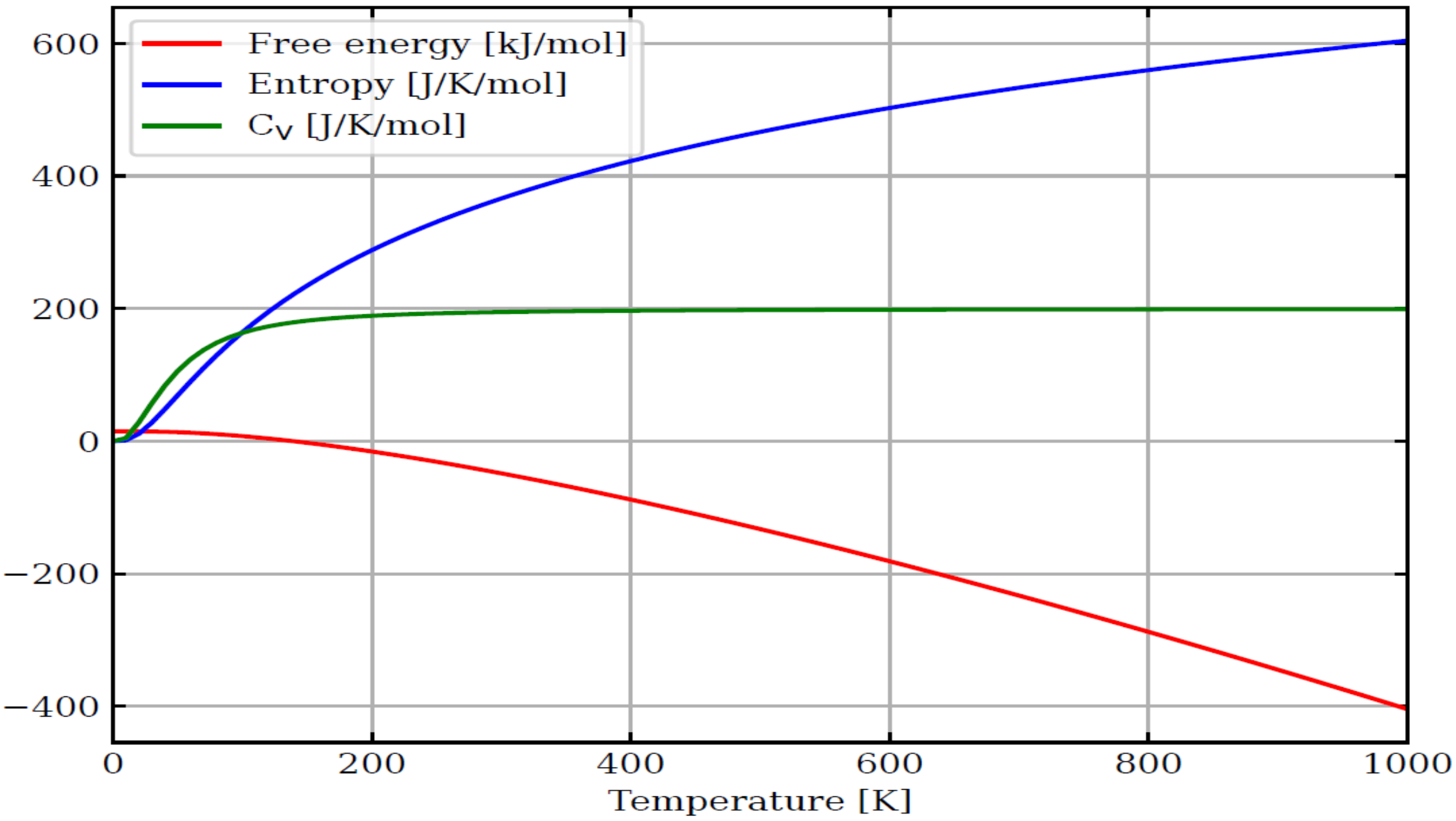

**Figure 8:** Temperature dependence of specific heat at constant volume $C_v$, free energy, and entropy with temperature.

The specific heat approaches Dulong-Petit value near room temperature. The entropy rises with an increase in temperature, which is the expected behavior. This is because increasing temperature increases lattice vibrations and the number of accessible microstates. Increase in temperature also reduces free energy, becoming negative in temperature near 200 K. This is the standard behavior of materials, increasing disorder or entropy causing decrease in free energy.

Thermoelectric materials are capable of directly converting heat energy into electrical energy. They can be useful in cogeneration systems [82]. We employed the BoltzTraP2 software package to compute various thermal properties of $Na_3Bi$ in order to evaluate its potential for thermoelectric power generation. BoltzTraP2 uses band structure data to calculate temperature dependent transport properties using the constant relaxation time approximation solving the linearized Boltzmann transport equation [30]. We have calculated several thermoelectric properties such as Seebeck coefficient ($S$), electronic thermal conductivity ($\kappa_e$), electrical conductivity ($\sigma$), power factor (PF), Figure of merit (ZT) for a range of temperature (200 K to 1000 K). The relaxation time is taken to be $\tau = 10^{-14}$ s, which is a typical value for electron scattering under constant relaxation time assumption in materials [83–86].

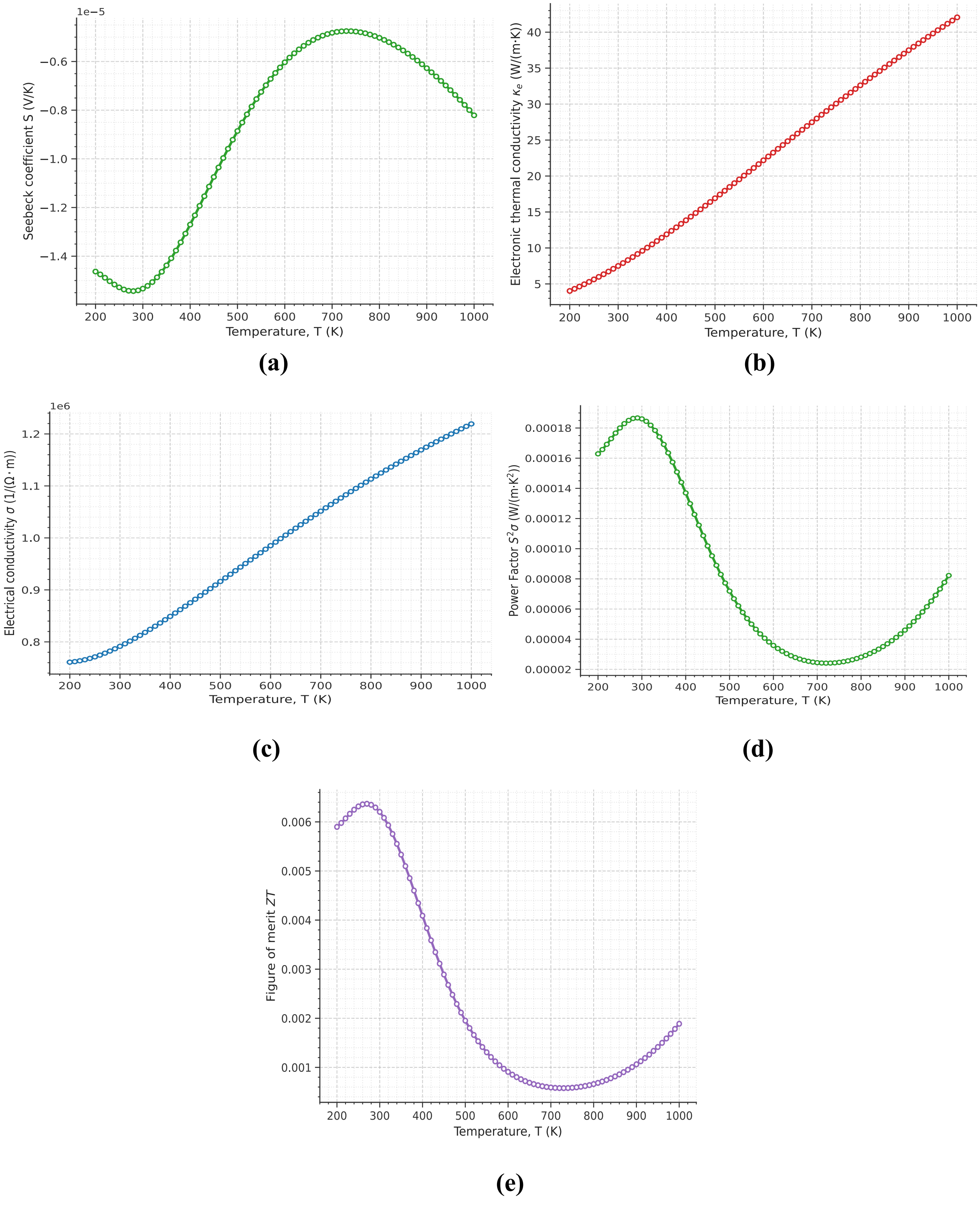

**Figure 9:** (a) Seebeck coefficient (b) electronic thermal conductivity (c) electrical conductivity (d) power factor (e) thermoelectric figure of merit for hexagonal phase of $Na_3Bi$.

Throughout the temperature range, the Seebeck coefficient is negative. This is an indication that electrons are the majority carriers, which is consistent with its semi metallic nature. The slight dip in 200K-300 K range increased scattering or reduced carrier mobility in that temperature range. Seebeck coefficient value has a peak near 750-800 K. This could indicate enhanced carrier diffusion or change in scattering mechanisms. Over 800 K, the drop in value might be an indication of bipolar conduction. Near 750 K, it is more useful in thermoelectric application. With increase in temperature, the electronic thermal conductivity and electrical conductivity both increase. Both of the conductivity values of $Na_3Bi$ are highest among previously studied Na-Bi compounds [15]. Increase in temperature will increase mobility of charge carriers, which will enhance both thermal and electrical conduction. Power factor PF($\mu W/m \cdot K^2$) peaks near 320 K**,** and then gradually decreases**,** reaching a minimum (~25 $\mu W/m \cdot K^2$) near 700 K**,** before slightly increasing again toward 1000 K. The peak near room temperature suggests suitability for early staged thermoelectric application. But enhanced electronic conductivity suggests this material in its pure form is not suitable for thermoelectric devices. We have used lattice thermal conductivity ($\kappa_l$) value to be 1.5 $WmK^{-1}$, which is reported for Bi based compounds [15]. We compared our calculated thermoelectric figure of merit (ZT) values with other Na-Bi compounds and cubic polymorph of $Na_3Bi$ (ranging from 0.10 to 0.53) [15]. Our study suggests that hexagonal polymorph of $Na_3Bi$ is not useful for thermoelectric application. Alloying, nano-structuring, introducing phonon scattering centers might improve its thermal power generation capability. This conclusion is consistent with previous reports, which suggest that effect of doping or pressure can improve this compound's thermoelectric performance [87,88].

### 3.10. Optical Properties

A material's optical properties characterize the ways in which it interacts with light. These properties play a significant role in characterizing the material and observing its possible applications in different optical devices. By analyzing the probabilities of photon-induced transitions between various electronic orbitals, the optical parameters presented in this section were determined.

We have calculated several energy-dependent optical characteristics for $Na_3Bi$ such as complex dielectric function ε(ω), refractive index, optical conductivity σ(ω), reflectivity R(ω), absorption coefficient α(ω), and loss function L(ω). These calculations were done for [100] and [001] electric field polarization directions and for incident photon energies from 0 to 20 eV.

While conducting the optical properties calculations, the complex dielectric function, $\varepsilon(\omega) = \varepsilon_1(\omega) + i\varepsilon_2(\omega)$, was evaluated first. The standard expression for the imaginary dielectric function $\varepsilon_2(\omega)$, within the independent-particle approximation is [89]:

$$\varepsilon_{\alpha\beta}^{(2)}(\omega) = \frac{4\pi^2 e^2}{\Omega} \lim_{q\to 0} \left(\frac{1}{q^2}\right) \sum_{c,v,k} \left(2\omega_k \delta(E_{ck} - E_{vk} - \omega) \langle u_{ck+e_\alpha q} | u_{vk} \rangle \left\langle u_{vk} \middle| u_{ck+e_\beta q} \right\rangle\right) \tag{35}$$

where, $\varepsilon_{\alpha\beta}$ is defined as: $D_\alpha(\omega) = \varepsilon_{\alpha\beta}(\omega) E_\beta(\omega)$.

$D_\alpha(\omega)$ is the αth component of displacement vector. $|u_\alpha\rangle$ is the Bloch state having momentum $\hbar\alpha$ .Using Kramers-Kronig relations, the real part of $\varepsilon$ is calculated [90].

The real and imaginary parts of complex dielectric function were used to derive the remaining optical properties, using established methodology [91].

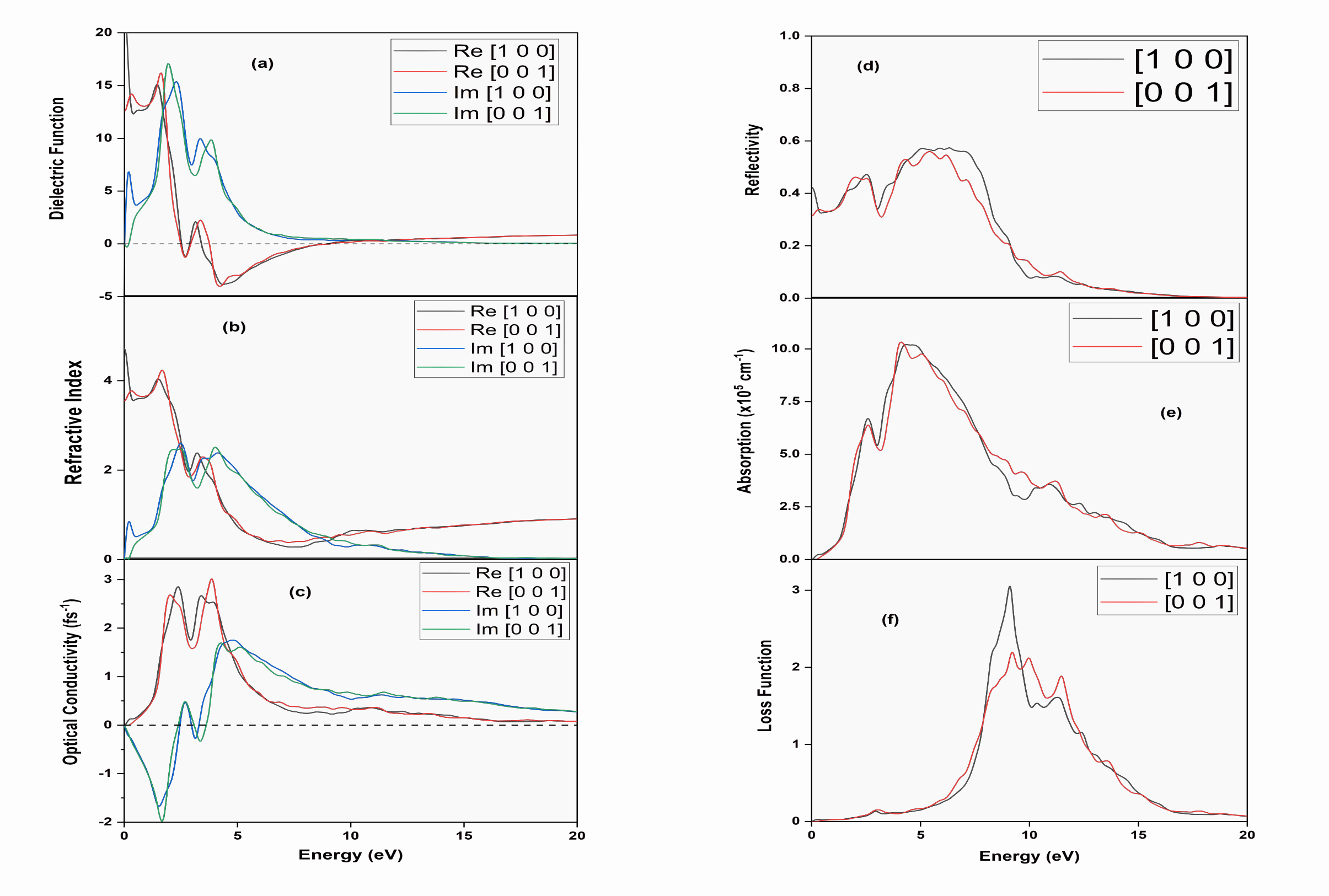

**Figure 10.** (a) Dielectric function (real & imaginary), (b) Refractive index (real & imaginary), (c) Optical conductivity (real & imaginary), (d) Reflectivity, (e) Absorption coefficient, and (f) Loss function of $Na_3Bi$, as a function of photon energy.

Figure 10(a) shows photon energy dependence of the real (Re) and imaginary (Im) parts of complex dielectric function ($\varepsilon(\omega)$). At low energies, negative value of real component indicates the metallic nature of $Na_3Bi$. Real part of dielectric function crosses zero near 9.5 eV, while imaginary part also remains negligible at that value. So, it is evident that the plasma frequency of $Na_3Bi$ is approximately 9.5 eV.

Figure 10(b) displays the real, $\eta(\omega)$, and the imaginary, $k(\omega)$, parts of the refractive index. The phase velocity of em wave in the material is associated with the real component of the refractive index. In visible region, the real component of the refractive index, $\eta(\omega)$, is relatively high, as seen in Figure 10(b). This suggests strong dielectric response in this material and also suggests that this material may be useful in optical confinement and waveguide related applications. The imaginary part, known as the extinction coefficient, quantifies the reduction in intensity as light passes through the material. Extinction coefficient, $k(\omega)$ becomes zero around the plasma frequency, indicating a substantially unobstructed propagation of electromagnetic radiation.

At zero photon energy, the optical conductivity $\sigma(\omega)$ has a nonzero finite value (Figure 10(c)). It is evident that $Na_3Bi$ has metallic nature. This conclusion is consistent with electronic properties calculation result in this work.

The reflectivity of $Na_3Bi$ is shown in Figure 10(d). It demonstrates that reflectivity, R(ω), drastically falls at the anticipated plasma frequency, or at 9.5 eV. Additionally, it can be observed from the reflectivity spectrum that the material is a moderate reflector of solar radiation in visible and UV spectra. Though experimental studies on reflectivity of pristine $Na_3Bi$ are lacking, an experimental study on n-doped $Na_3Bi$ reveal almost similar characteristic of reflectivity spectra [16]. These results suggest its potential applications in optoelectronic components, where controlled reflection over a broad spectral range is desirable [92]. However, further device-specific studies are required before any definitive claims regarding its practical applicability can be made.

In Figure 10(e), the absorption coefficient of $Na_3Bi$ is shown. The absorption coefficient of a material is a key factor in determining its effectiveness as a photon absorber. The non-zero absorption at 0 eV is a reassuring indicator of the metallic nature of $Na_3Bi$. The absorption is significantly high in visible and UV region ($10^5$–$10^6$ $cm^{-1}$). Materials with strong absorption coefficients in UV region are suitable for UV absorbers, like ZnO, $TiO_2$, $CeO_2$, Graphene Oxide etc [93–96]. So, our results suggest strong light matter interaction in visible and UV region, makes it promising candidate for UV photodetection and related optoelectronic applications [47,56,97,98]. Additionally, for solar coating devices, materials with very high absorbance (0.84 ~ 0.95) are useful [99,100]. However, experimental validation is required to confirm the suitability of $Na_3Bi$ in these optoelectronic applications.

Loss function characterizes the energy dissipation of electron moving through the material. The maximum value of loss function corresponds to the plasma frequency. The maximum value of the loss function, L(ω) for $Na_3Bi$ is approximately at 9.5 eV, as shown in Figure 10(f). This finding is in agreement with the spectrum characteristics of all other optical parameters. The loss function peak corresponds to resonance frequency of collective vibration of free electrons in a metallic system [70,101].

The optical constants show minimal anisotropy with respect to the orientation of the photon polarization. For instance, the conductivity, reflectivity, and loss function spectrum exhibit subtle variations along the [100] and [001] orientations.

## 4. Conclusions

This work presents a comprehensive first-principles characterization of hexagonal $Na_3Bi$, addressing several elasto-mechanical, thermophysical, and thermoelectric properties that had not been previously reported for this phase. The material is mechanically and dynamically stable, with a brittle, predominantly covalent bonding character supplemented by a modest metallic contribution; its mechanical response is governed more strongly by shear than by unidirectional strain, and its subtle elastic and optical anisotropy reflects the underlying low bonding anisotropy of the lattice. Thermophysically, the low Debye temperature and acoustic impedance indicate a soft, loosely bound structure well suited to sound-damping applications, while the calculated power factor and figure of merit suggest that pristine $Na_3Bi$ is not, on its own, an efficient thermoelectric material — though doping, alloying, or nanostructuring may improve this.

Electronically, our band structure, density of states, Fermi surface, and charge density analyses jointly confirm $Na_3Bi$'s identity as a robust three-dimensional Dirac semimetal, with well-defined Dirac points pinned at the Fermi level and mixed covalent-metallic bonding consistent with its mechanical behavior. Optically, the material combines a high visible-range refractive index, moderate broad-spectrum reflectivity, and strong UV absorption, with only weak dependence on photon polarization direction — properties that point toward potential use in UV photodetection and optical confinement applications. Taken together, these results establish a broad physical baseline for hexagonal $Na_3Bi$ and motivate further experimental validation, particularly of its predicted acoustic, thermoelectric, and optoelectronic behavior.

**Acknowledgements**

We are thankful to Bangladesh Research and Education Network (BdREN) for providing access to their server, which significantly facilitated our calculations using the Vienna Ab initio Simulation Package (VASP).

**Data availability statement**

The data that support the findings of this study are available from the corresponding author, IMS, upon reasonable request.

**Declaration of interests**

The authors declare that they have no known competing financial interests or personal relationships that could have appeared to influence the work reported in this paper.

**Funding Declaration**

The author(s) received no financial support for the research, authorship, and/or publication of this article.